\documentstyle[preprint,aps]{revtex}
\begin{document}
\draft
\title{Fractional Quantization and Fractional Quantum Hall Effect}
\author{Hyeong Rag Lee}
\address{Department of Physics, Kyungpook National University,
  Taegu, 702-701 Korea}
\date{\today}
\maketitle

\begin{abstract}
  We present a fractional quantization in a two dimensional space.
  The angular momenta of the two dimensional electrons are quantized
  in fractional numbers by the boundary conditions on a multi-layered
  Riemann surface. Extended wave functions for the incompressible
  quantum fluid states are presented and the cohesive and the excitation
  energies are given.
\end{abstract}
\vspace{0.1in}
\vspace{0.3in}

 Since the experimental observation of the Fractional Quantum Hall
 Effect(FOHE)[1] there have been a variety of theoretical approaches
 as well as  experimental works[2,3] to understand it.
 This remarkable phenomenon occurs when a two dimensional
 electron system is under a strong magnetic field in the quantum limit
 $\omega_c \tau >> 1$ where $\omega_c = \frac {eB_0}{m_e c}$
 is the cyclotron frequency and $\tau$ is the electronic scattering time.
 Hall conductance shows plateaus at certain values of the rational filling
 factor $\nu = \frac{n}{m}$[2,3], where n and m are integers with m being
 odd, i. e. $\sigma_{xy} = \frac {\nu e^2}{h}$. The filling factors for
 $\nu = \frac {1}{m}$ are associated with the formation of a uniform
 incompressible quantum fluid state so called Laughlin liquid[4].
 The quasi-particles of charge $\frac{e}{m}$ are responsible for the
 formation of such liquid. The hierarchy states for the general fillings
 are proposed with the scheme of the quasi-particles[5] and the
 composite fermion approach[6] long time ago.
 However, the ground states and the excited
 states for the general fillings are not clear yet.
 In this paper, we provide the ground state wave functions for an
 arbitrary filling($\nu = \frac{n}{m}$)
 and the excited states associated with the fractionally
 charged quasi-particles $\frac{e}{m}$ on a Riemann surface .

  It is clear that the statistics in two dimensional space is different
  from those in three dimensions, because the former space is represented
  by the Braid group[7] while the latter is represented by the permutation
  group[8]. The permutation group allows only two possible statistics
  bosonic and fermionic statistics. However fractional statistics[5,6,9,10]
  is allowed in the context of the Braid group representation. The
  basic generator of the Braid group is the exchange of two particles
  in the configuration space. Consequently, if we consider the interchange
  of the positions of two identical particles in two dimensions, the wave
  function obtains a phase factor,
\begin{equation}
 \Psi \left( z_2, z_1 \right) = e^{i \theta} \Psi \left( z_1, ~z_2 \right).
\end{equation}
  Here $\frac {\theta}{\pi}$ can be any real number. (i) When
  $\frac {\theta}{\pi}$ is an even integer or 0, the particles are bosons.
  (ii) When $\frac {\theta}{\pi}$ is an odd integer, the particles are
  fermions. (iii) For other $\frac {\theta}{\pi}$ real numbers, the
  particles are called anyons[10]. When $\frac {\theta}{\pi} = \frac {m}{n}$,
  where m and n are arbitrary integers, the particles are quantized in
  fractional numbers by the following arguments. Wave functions for the
  fractional angular momenta are multi-valued functions of the positions
  in the multiply connected fundamental space except for the branch points
  which are located at 0 and $\infty$. A Riemann surface for the
  fractional quantization is obtained by replacing the two dimensional
  plane with a surface made up of n sheets
  $R_0 , R_1 , \cdots  , R_n$, each cut along the positive real axis with
  the common origin. The lower edge of the slit in the first sheet is
  joined to the upper edge of the slit in the second sheet with the
  exchange of two branch points(0, $\infty$), which flips the second
  sheet and makes the rotation in the same direction. The lower edge of
  the slit in the second sheet is joined to the upper edge of the slit
  in the third sheet in the same manner and so on until the last sheet.
  By joining the upper edge of the slit in the last sheet to the lower
  edge of the slit in the first sheet with the exchange of two branch
  points, we can construct an extended Riemann surface  which is closed
  and simply connected. The geometric device for n=3 is shown in figure 1.
  The wave function is a continuous single-valued function of complex
  variables on the extended Riemann surface with $\theta = [0, 2n \pi]$.
  If we continuously interchange the positions of two particles 2n times
  in the same direction, which corresponds to winding n times of particle 1
  around particle 2, the wave function will change by a complex phase factor
  $e^{\theta(2n \pi)} = e^{2m \pi} = 1$.
  By the boundary condition of joining the wave smoothly on itself after
  the interchange of the particles 2n times the wave function is quantized
  on the extended Riemann surface. The angular momentum is quantized in
  integral numbers m on the n layered Riemann surface, which is associated
  with the fractional angular momentum $\frac {m}{n}$ in the projected two
  dimensional space. In the case after particle 1 interchanges n times with
  particle 2 in the same sense, so that 1 ends where 2 began and vice versa,
  the phase change $e^{\theta (n \pi)} = e^{m \pi}$. For an odd integral m,
  the particles are fermions.

  We consider the electrons confined to the x-y plane under a transverse
  magnetic field $B_0 \hat z$. Ignoring the electron-electron interaction,
  two electron Hamiltonian is given by
\begin{equation}
   H = \sum_{j=1}^{2} \frac{1}{2m_e} |\frac{\hbar}{i} \nabla_j - \frac{e}{c} {\bf  A}_j|^2,
\end{equation}
  where ${\bf A} = \frac{1}{2} B_0 (y \hat x -x \hat y )$ is the symmetric
  gauge vector potential. This problem separates into the center of mass
  coordinates and the relative coordinates given by,
\begin{equation}
Z = \frac {(z_1 + z_2 )}{\sqrt 2}, z = \frac {(z_1 - z_2 )}{\sqrt 2},
\end{equation}
  where $z_j = x_j - i y_j$ is a complex number locating the j-th electron.
  The center of mass motion is trivial and quantized in the ordinary Landau
  levels with the Hamiltonian of the form
\begin{equation}
H_{cm} = \frac{1}{2m_e} | \frac{\hbar}{i} \nabla -\frac{e}{c} {\bf A}|^2,
\end{equation}
  where the mass is $m_e$ at the center of mass in this transformation. The Hamiltonian
  for the relative motion is given as
\begin{equation}
H_{rel} = \frac{1}{2m_e} | \frac{\hbar}{i} \nabla -\frac{e}{c} {\bf A}|^2,
\end{equation}
  in two dimensional configuration space.
  Using the dimensionless units [energy in units of cyclotron energy
  $\hbar \omega_c = \hbar \frac{eB_0}{m_e c}$, length in units
  of magnetic length $a_0 = \left( \frac{\hbar}{m_e \omega_c} \right)^{1/2}$], we obtain
\begin{equation}
\frac{H_{rel}}{\hbar\omega_c} = -\frac{1}{2}\nabla^2 + \frac{i}{2}
 \frac{\partial}{\partial\theta} + \frac{1}{8}\rho^2,
\end{equation}
where $\rho = (x^2 + y^2)^{\frac{1}{2}}$. For the internal motion,
  as discussed earlier of this paper, fractional quantization
  $\frac{\theta}{\pi} = \frac{m}{n}$
  is allowed in a two dimensional space. The magnetic flux seen by an electron is
  $e\oint {\bf A} \cdot d {\bf r}$ in the two dimensional configuration space.
  To achieve the same magnetic flux on an n layered Riemann
  surface the electron separates into n quasi-particles of the reduced
  charge $e^* = \frac{e}{n}$ and the reduced mass $m_e^* =\frac{m_e}{n}$
  in each plane. The n quasi-particles in each layer are sticked together
  with the common origin by the property of the Rieman surface.
  Therefore the effective Hamiltonian for the relative coordinate is written by
\begin{equation}
H_{rel}^{eff} = \frac {1}{m_e^*} | \frac{\hbar}{i} \nabla - \frac{e^*}{c} {\bf A} |^2,
\end{equation}
  on the Riemann surface. We adopt energy and length scales in which the
  cyclotron energy
\begin{equation}
\hbar \omega_c = \hbar \frac{e^* B_0}{m_e^* c} = \hbar \frac{eB_0}{m_e c},
\end{equation}
  and the magnetic length
\begin{equation}
a_n = \left( \frac{\hbar}{m_e^* \omega_c} \right)^{1/2} = \sqrt n a_0.
\end{equation}
  In dimensionless units [energy in units of the same cyclotron energy $\hbar \omega_c$,
  length in units of magnetic length $a_n$], the effective Hamiltonian is transformed into
\begin{equation}
\frac{H_{rel}^{eff}}{\hbar\omega_c} = -\frac{1}{2}\nabla^2 + \frac{i}{2}
 \frac{\partial}{\partial\theta} + \frac{1}{8}\rho^2,
\end{equation}
 which has exactly the same form as given in eq. (6). Here the angle
 $\theta$ can vary from 0 to $2n\pi$ on the Riemann surface of
 n sheets. The Landau level wave functions are polynomials in $\rho$ times a
 Gaussian function. We are therefore led to try a power series in $\rho$ times a
 Gaussian function as the solution. The requirement that the power series
 must terminate gives the following energy eigenvalues:
\begin{equation}
E_{rel} = \hbar\omega_c \left( n_{rel} + \frac{1}{2}|l_{rel}| + \frac{1}{2}l_{rel}
  +\frac{1}{2} \right),
\end{equation}
 where $n_{rel}$ is positive integer and $l_{rel}$ is the angular momentum of the quasi-particles.
 The angular momentum $l_{rel}$ of a quasi-particle has a fractional value
 $\frac{m}{n}$ in each layer, that is, each quasi-particle is quantized
 in fractional numbers with $\theta=[0, 2\pi]$. The angular momentum for the n
 degenerate quasi-particles which are sticked together on the Riemann surface will be
 $m$ in the expanded space of $\theta=[0, 2n\pi]$. The value of $m$ is even integral
 for a symmetric state and odd integral for an antisymmetric state. Therefore,
 the eigen energy spectrum represents just the ordinary Landau levels.
 If we consider the lowest Landau level, the eigen state for the internal motion
 can be obtained as
\begin{equation}
\phi_{\frac{m}{n}} = \frac{1}{[2n\pi 2^{\frac{m}{n}} \Gamma(\frac{m}{n})]^{1/2}} z^{\frac{m}{n}} e^{-\frac{1}{4}|z|^2},
\end{equation}
  where the value of n in the normalization constant is due to the expansion of the
  angle $\theta$ from $[0,2\pi]$ to $[0,2n\pi]$ on the Rieman surface.
  The cyclotron motion in the relative coordinates has a fractional angular momentum
  $\frac{m}{n}$ about the origin in two dimensional space. These states (with many different
  values of m and n) are constructed from the statistics of the identical particles in
  two space dimensions and can not be derivable from any single particle state.
  These many particle states can now be utilized in two dimensional systems
  with a proper thermodynamic limit. For the fractional quantum Hall effect at
  a filling less than one, the wave function for the relative motion should be
  antisymmetric and m becomes odd integral in this case.

  The model Hamiltonian for the N electrons confined in two dimensions under a
  transverse magnetic field can be written
\begin{equation}
   H = \sum_{j=1}^{N} \frac{1}{2m_e} |\frac{\hbar}{i} \nabla_j - \frac{e}{c} {\bf A}_j|^2
      + \sum _{j<k} V_{e-e}(|{\bf r}_j - {\bf r}_k|) + \sum_j V_{b-e}({\bf r}_j),
\end{equation}
  where $V_{e-e}$ is the Coulomb repulsion between electrons and $V_{b-e}$ is the
  one body background potential due to a uniform density of positive charge.
  In the limit of high magnetic field, the electrons are at the lowest Landau level.
  As a result, the first term is simply a constant. For a short range interaction
  (shorter than $r^{-2}$ as $r\rightarrow \infty$), the background potential is
  simply a constant, except close to the edges of the sample[11]. Fractional qunatum
  Hall states arise from a condensation of the two dimensional electrons into
  a collective state, i.e., incompressible quantum fluid state, as a result
  of repulsive interelectron interactions[4].

  We can thus construct the extended wave functions for the incompressible
  quantum fluid of the form[4]
\begin{equation}
\Psi_\mu (z_1 , \cdots ,z_N ) =  \prod_{j>k}^N (z_j - z_k )^\mu exp \left( - \frac{1}{4}
\sum_l | z_l| ^2 \right),
\end{equation}
  where $\mu = \frac{m}{n}$. The states $\Psi_\mu$ are translationally invariant
  and eigenstates of total angular momentum. Since the total angular momentum
  $M$ is a good quantum number in the projected two dimensional configuration space,
  we use the conventional definition of the filling factor $\nu=N(N-1)/2M$.
  Therefore, the wave functions $\Psi_\mu$ represent the hierarchy states
  for the general filling factor $\nu = \frac{n}{m}$ in FQHE.
  The electronic charges separate into the n degenerate layers of the Rieman surface.
  In each layer the quasi-particles of  fundamental charge $\frac{e}{m}$ condensate into
  the Laughlin states. If n such degenrate layers are
  projected to the two dimensional space, the composite particles of charge
  $\frac{n}{m}e$ condensate into the incompressible quantum fluid states with
  a filling factor $\frac{n}{m}$.
  This description for the composite particle is the same as the
  degenerate Landau levels for the quasi-particles in the composite
  fermion theory propoesd by Jain[6].
  For n = 1, we can recover Laughlin's results of filling factor $\frac{1}{m}$.

The ground state energy of $\Psi_\mu$ can be obtained by the ordinary
  hypernetted chain approximation[3,4]
\begin{equation}
U_{total} \left( \frac{m}{n} \right) =  \frac{0.814}{\sqrt{m}} \left[\frac{0.230}{(m/n)^{0.64}}
-1 \right] \frac{e^2}{a_0}$$.
\end{equation}
  The cohesive energy of the incompressible quantum fluid decreases as m increases.
  For the given m, the cohesive energy decreases as n increases:
  the factor $n^{0.64}$ in the first term of (15) reflects the larger magnetic
  length for the quasi-particles in n-layered structure.

 We define the elementary excitations of $\Psi_\mu$ as
\begin{equation}
\Psi_\mu^{(z_0 )} = \prod_i^N (z_i - z_0 )^{\frac{1}{n}} \Psi_\mu.
\end{equation}
  Writing $|\Psi_\mu ^{(z_0 )}|^2 = e^{-H_\mu (Z_0 )}$,
\begin{equation}
H_\mu (z_0 ) = -2 \mu \sum_{j<k}^N ln |  z_j - z_k | + \frac{1}{2} \sum_l |  z_l |^2 +
\frac{2}{n} \sum_i ln | z_i - z_0 |.
\end{equation}
  The elementary excitations of $\Psi_\mu$ are particles of charge  $\frac{1}{m}$ by
the Berry phase calculation[12]. We can calculate the excitation energy to make the quasihole using
  the two-component hypernetted chain approximation[3,4]. The results for
  $\left( \frac{n}{m} \right)$ $= \left( \frac{1}{3}, \frac{2}{3} \right)$;
  $( \frac{1}{5}$, $\frac{2}{5}, \frac{3}{5}$, $\frac{4}{5} )$;
  and $( \frac{1}{7}, \frac{2}{7}$, $\frac{3}{7}, \frac{4}{7}, \frac{5}{7}$, $\frac{6}{7} )$
  are $(0.0095, 0.0079) \frac{e^2}{a_0}$,
  $( 0.0018, 0.0015, 0.0013$, $0.0012) \frac{e^2}{a_0}$ and
  $( 0.00059, 0.00050, 0.00045$, $0.00041, 0.00038$, $0.00037) \frac{e^2}{a_0}$
  respectively[13]. These results are close to the experimental results
  observed[14].

  In conclusion, the angular momenta of the two dimensional electrons are
  quantized in fractional numbers on a multi-layered Riemann surface. The
  incompressible quantum fluid of the filling factor $\nu = \frac{n}{m}$
  is related with these fractional quantum numbers. The cohesive energy of the
  two dimensional electron system is
  reduced by the expansion of the inter-electronic distance on a multi-layered
  Riemann surface. The effect of disorder[15] in the system is also reduced on
  the multi-layered Riemann surface. As a result, Hall plateau is observed for
  bigger values of n as m becomes large in the experiment.
  We believe the Hall plateau can be observed for smaller values
  of n with the same m observed in the experiment if the sample is cleaned
  to a disorder free system.

  This research is partially supported by the Korea Ministry of Education
  (BSRI 97-2404) at Kyungpook National University.

\vspace{0.2in}
Electronic Address: phyhrlee@bh.kyungpook.ac.kr

\begin{center}
Figure Caption
\end{center}
\begin{description}
\item[Fig. 1.]
  Fig. (1). Geometry of three layered Riemann surface.
\end{description}

\begin{references}
\bibitem{Tsui} D. C. Tsui, H. L. Stormer, and A. C. Gossard,  Phys. Rev. Lett. {\bf 48},
   1559 (1982)
\bibitem{Prange} {\it The  Quantum Hall  Effect}, edited  by R.  E. Prange  and S.  M.
   Girvin (Springer-Verlag, New York, 1990), and references therein.
\bibitem{Chakraborty} For the latest review see:  T. Chakraborty and P. Pietilainen,  {\it
   The Quantum Hall Effects  : Fractional and Integral},  2nd ed. (Springer-Verlag, Berlin,
   1995), and references therein.
\bibitem{Laughlin} R. B. Laughlin, Phys. Rev. Lett. {\bf 50}, 1395 (1983); R. B. Laughlin,
   Surf. Sci. {\bf 142}, 163 (1984)
\bibitem{Haldane} F. D. M. Haldane, Phys. Rev. Lett. {\bf 51}, 605(1983);
   B. I. Halperin, Phys. Rev. Lett. {\bf 52}, 1583(1984);
\bibitem{Jane} J. K. Jain, Phys. Rev. Lett. {\bf 63}, 199(1989);
   J. K. Jain, Phys. Rev. B {\bf 41}, 7653(1990)
\bibitem{Birman} Cf. J. Birman, {\it Braids,  Links, and Mapping Class Groups},  Annals
   of Math. Studies 82 (Princeton University Press, Princeton, 1973)
\bibitem{Lee} H. C. Lee, M. L. Ge, and M. Couture, Int. Jour. Mod. Phys. {\bf A4}, 2333
   (1989); J. M. Leinaas, and J. Myrheim, Nuo. Cim. {\bf 37B}, 1 (1997)
\bibitem{Sect.}  See Sect. 6.7 in Ref. 3 and references therein.
\bibitem{Wilczek} F. Wilczek, Phys. Rev. Lett. {\bf 49}, 957(1982)
\bibitem{Trugman} S. A. Trugman and S. Kivelson, Phys. Rev. B {\bf 31}, 5280(1985)
\bibitem{Arovas} D. Arovas, J. R. Schrieffer, F. Wilczek, Phys. Rev. Lett. {\bf 53}, 722(1984)
\bibitem{} In the two-component hypernetted chain  approximation the quasihole density
   is not  properly given  in Refs.  3 and   4. Since the  elementary excitation  of charge
   $\frac{e}{m}$
   occupies the same area of an electron, the density of a quasihole $\frac{e}{m}$
   is reduced by a factor $\frac{1}{m}$
   compared to the electron density.
\bibitem{Sect.} See Sect. 6.6 in Ref. 3. and references therein.
\bibitem{Gold} A. Gold, Europhys. Lett. {\bf  1}, 241, 479(E), (1986); R. B.  Laughlin, M.
   L. Cohen, J. M. Kosterlitz, H. Levine,  S. B. Libby, A. M. M. Pruisken,  Phys. Rev. {\bf
   B32}, 1311 (1985); R. B. Laughlin, Surf. Sci. {\bf 170}, 167 (1986)

\end{references}
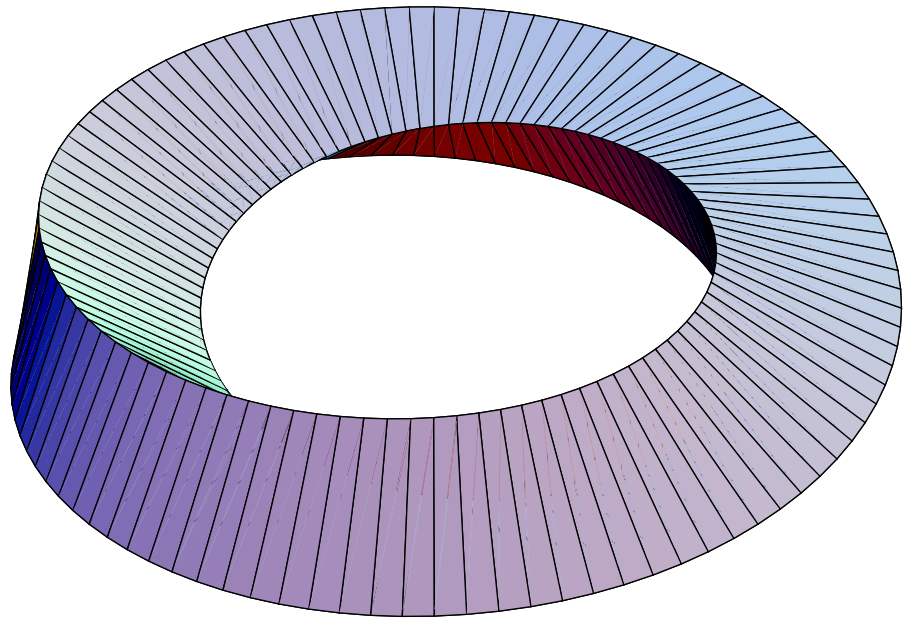
\end{document}